# Pre-selectable integer quantum conductance of electrochemically fabricated silver point contacts

F.-Q. Xie[1,4]*, R. Maul[2]*, S. Brendelberger[1], Ch. Obermair[1,4], E.B. Starikov[2], W. Wenzel[2,4], G. Schön[2,3,4], Th. Schimmel[1,2,4]

[1] Institut für Angewandte Physik, Universität Karlsruhe, 76128 Karlsruhe, Germany
[2] Forschungszentrum Karlsruhe, Institut für Nanotechnologie, 76021 Karlsruhe, Germany
[3] Institut für Theoretische Festkörperphysik, Universität Karlsruhe, 76128 Karlsruhe, Germany
[4] DFG-Center for Functional Nanostructures (CFN), Universität Karlsruhe, 76128 Karlsruhe, Germany

* authors contributed equally

The controlled fabrication of well-ordered atomic-scale metallic contacts is of great interest: it is expected that the experimentally observed high percentage of point contacts with a conductance at non-integer multiples of the conductance quantum $G_0=2e^2/h$ in simple metals is correlated to defects resulting from the fabrication process. Here we demonstrate a combined electrochemical deposition and annealing method which allows the controlled fabrication of point contacts with pre-selectable integer quantum conductance. The resulting conductance measurements on silver point contacts are compared with tight-binding-like conductance calculations of modeled idealized junction geometries between two silver crystals with a predefined number of contact atoms.

Due to their interesting physical properties and potential technological perspectives, metallic quantum wires and atomic-scale contacts are an object of intensive experimental[1-9] and theoretical investigations[10-13]. As the size of these constrictions is smaller than the scattering length of the conduction electrons, transport through such contacts is ballistic, and as the width of the contacts is on the length scale of the electron wavelength, the quantum nature of the electrons is directly observable. The electrical conductance of such quantum structures is given by the Landauer formula $G = G_0 \Sigma \tau_n$, where $G_0 = 2e^2/h$ is the conductance quantum (where $e$ is the charge of the electron and $h$ is Planck's constant). $\tau_n$ gives the transmission probability of the $n$th channel and depends crucially on the orbital structure of the conducting atoms[3], but also on the atomic structure, in particular on scattering at defects and boundaries[11], as well as internal stress[10]. In practice, most conductance measurements of point contacts, even for simple metals, yield non-integer multiples of the conductance quantum $G_0$. Such deviations from the ideal behavior can stem from material-specific properties of the junction or from defects that result from the fabrication process. Here we combine electrochemical deposition and annealing methods for the fabrication of metallic quantum point contacts which yield nearly ideal integer multiples of $G_0$ for the quantum conductance and explain their properties by comparison with conductance calculations for selected, near crystalline junction geometries with a pre-selected number of contact atoms.

Especially in experiments based on atomic-scale contact fabrication by mechanical deformation (e.g. break junctions or STM setups[1-3]), there is very limited control of the growth and properties of the atomic-scale contacts. In these experiments long-term stable and defect-free contacts with conductance at integer multiples of the conductance quantum $G_0$ are difficult to realize in practice as the fabrication process is essentially connected with the formation of atomic-scale defects such as dislocations.

To produce well-ordered contacts, a technique of nearly defect-free growth by slow quasi-equilibrium deposition is required, which can be provided by electrochemical deposition methods[4-8]. In addition, techniques of electrochemical annealing provide the possibility of healing atomic-scale defects in contacts even after fabrication (see below). Due to its high electrochemical exchange current density[14], silver is a promising candidate for efficiently applying electrochemical annealing techniques.

In this letter we demonstrate an electrochemical annealing method by electrochemical deposition/dissolution cycling of atomic-scale silver contacts and compare the experimentally observed conductance with the calculated conductance of modeled idealized junctions between two ideal single crystals with a predefined number of contacting atoms.

The experimental set-up is shown in Fig. 1(a). By applying an electrochemical potential, silver was deposited within the gap between two macroscopic gold electrodes (gap width typically of the order of 50 nm). The gold electrodes (thickness approx. 100 nm) were covered with an insulting polymer coating except for the immediate contact area and served as electrochemical working electrodes. Two silver wires (0.25 mm in diameter, 99.9985% purity) served as counter and quasi-reference electrodes. The potentials of the working electrodes with respect to the quasi-reference and counter electrodes were set by a computer-controlled bipotentiostat. The electrolyte consisted of 1 mM $AgNO_3$ + 0.1 M $HNO_3$ in bi-distilled water. All experiments were performed at room temperature, the electrolyte being kept in ambient air. For conductance measurements, an additional voltage of 12.9 mV was applied between the two gold electrodes. While one of the gold electrodes was connected to the ground potential the other gold electrode was kept at -12.9 mV relative to this ground potential.

When applying an electrochemical potential of 10 … 40 mV between the electrochemical reference electrode and the two gold electrodes (gold electrodes with negative bias relative to the electrochemical reference electrode), silver crystals formed on the two gold electrodes, two crystals finally meeting each other by forming an atomic-scale contact (see inset in Fig. 1(a)). During deposition, the conductance between the two gold electrodes was continuously measured. As soon as a predefined conductance value was exceeded, the computer-controlled feedback immediately stopped further deposition of silver on the working electrodes. If desired, the deposited contact could be fully or partially electrochemically dissolved by applying an electrochemical potential of -15 … -40 mV.

Figure 1(b) gives conductance-vs-time curves of the closing processes of four different atomic-scale contacts during initial deposition, i.e. before electrochemical annealing. In this way, initially, contacts of limited stability were formed, typically exhibiting conductance values which are *non*-integer multiples of $G_0$. Now, a dissolution/deposition cycle between predefined conductance values was performed: after the initial deposition cycle, a dissolution potential was applied until the conductance dropped below a predefined lower threshold. Subsequently, deposition was started once more until conductance exceeded a predefined upper threshold. At this point, a new dissolution-deposition cycle was started and so on. Typically, after a number of cycles, a stable contact was formed, which exhibited an integer conductance value, and the cycling was stopped. Using this method, stable conductance levels at integer multiples of $G_0$ were configured. Examples for $n \cdot G_0$ (n = 1, 2, 3, 4, 5) are given in Fig. 2(a). This transition from instable contacts with non-integer conductance to stable contacts with integer conductance values can be explained by an electrochemical annealing process, which heals defects in the direct contact region by electrochemical deposition and dissolution leading to an optimized contact configuration. After the electrochemical annealing process, most transitions appear to be instantaneous within the time resolution of the diagram of Fig. 2(a) (50 ms), whereas at higher time resolution (10 μs), fingerprints of the atomic-scale reorganization of the contact were observed in the form of both integer and non-integer instable transient levels.

In order to get insights into the possible structures of the measured point contacts, we calculated the coherent conductance of ideal crystalline silver nanojunctions (see Fig. 2(b)). Geometries were generated by assuming two fcc electrode clusters, which are connected at their tips by a small number of Ag-Ag-bridges in [111] direction with a bond length of 2.88 Å[15].

The zero-bias quantum conductance of a given junction geometry was computed with the Landauer formula[16,17]. The electronic structure was described using an extended Hückel model[18,19] including s-, p- and d-orbitals for each silver atom (around 3600 orbitals per junction). Consistently, material-specific surface Green's functions were computed using a decimation technique[17]. To reduce the influence of interference effects, we averaged the conductance $G(E)$ over a small interval [$E_F - \Delta$, $E_F + \Delta$] around the Fermi energy (with $\Delta$ = 50 meV), which is comparable to the temperature smearing in measurements at room-temperature.

As shown in Fig. 2(b), we find nearly integer conductance of the idealized geometries for contact geometries #1 … #5: 0.97 $G_0$, 1.95 $G_0$, 2.89 $G_0$, 3.95 $G_0$, 4.91 $G_0$, respectively. The deviation from integer multiples of $G_0$ of about 0.1 $G_0$ is within the range of the accuracy of our numerical method. We observe a good correlation between the number of silver atoms at the point of minimal cross section and the number of conductance quanta, which aids in the construction of geometries with a particular value of the conductance. However, this is a material-specific property of silver not necessarily to be encountered in other materials.

Figure 3 shows the calculated transmission as a function of the electron energy within the energy interval [$E_F$ - 6 eV, $E_F$ + 6 eV] for the five silver point contact geometries (#1 - #5) given in Fig. 2(b). The experimentally relevant values correspond to the conductance at the Fermi energy indicated by the vertical line in the figure. For the given silver junction geometries we obtained Fermi energies between -5.83 eV and -5.81 eV, which may be slightly below the correct value, caused by the known energy underestimation of the extended Hückel model[17]. The conductance curve oscillations are sensitive to the atomic positions. Therefore, an average of the conductance around the Fermi energy yields a more representative value of the conductance $G$, taking effectively into account the atomic vibrations during the measurement.

In order to study to which extent the conductance values change due to geometrical changes in the interatomic distance of the contacting atoms and the relative angle between the contacting crystals, we introduced finite changes in contact geometry: we calculated the electrode distance and twist-angle dependence of the zero bias conductance. Increasing the electrode distance to twice the Ag-Ag bond length leads to a decrease by 86.7 % in the conductance, while twisting the electrodes by 60° against each other leads to a decrease of conductance of 22.0 %.

To conclude, the results demonstrate that for silver as a representative of a simple s-type metal, if defects and disorder in the contact area are avoided, the conductance in atomic-scale point contacts typically is an integer multiple of the conductance quantum $G_0$. The method of combined electrochemical deposition and electrochemical annealing of point contacts has proven to be a very efficient technique to generate such well-ordered contacts. On the other hand, if annealing is omitted, non-integer multiples of the conductance quantum are observed, which can be attributed to scattering due to defects and disorder within the contact area. These observations are confirmed by calculations on ideal model geometries of contacting silver nanocrystals, which yield integer multiples of the conductance quantum within the accuracy of the calculation in all five cases investigated. As soon as disorder or local distortions of the atomic lattice within the contact area are introduced in the model geometry, drastic deviations from integer quantum conductance are obtained. This, in turn, indicates that such kind of disorder is effectively avoided in our experiments as a consequence of the electrochemical annealing approach. The results not only give an experimental proof of integer conductance quantization in annealed contact geometries of simple metals. The reproducible fabrication process also opens perspectives for the controlled configuration of atomic-scale quantum devices.

This work was supported by the Deutsche Forschungsgemeinschaft within the Center for Functional Nanostructures (CFN), project B2.3 as well as grant WE 1863/15-1. The SEM image was taken at the Laboratory for Electron Microscopy (LEM) of the Universität Karlsruhe. We acknowledge the use of the computational facilities at the Computational Science Center at KIST, Seoul.

Figure captions

FIG. 1.

(a) Schematic diagram of the experimental setup. Within a narrow gap between two gold electrodes on a glass substrate, a silver point contact is deposited electrochemically. Inset: Two electrochemically deposited silver crystals between which the atomic-scale silver contact forms (deposition voltage: 30 mV).

(b) Conductance of four different silver point contacts during initial electrochemical deposition. *Before* electrochemical annealing, contacts of limited stability are formed, typically exhibiting conductance values which are *non*-integer multiples of $G_0$.

FIG. 2.

Comparison of experimental conductance data of electrochemically annealed silver point contacts with calculations assuming idealized geometries.

(a) Quantum conductance of five different annealed atomic-scale contacts at 1 $G_0$, 2 $G_0$, 3 $G_0$, 4 $G_0$, 5 $G_0$, respectively (with 1 $G_0 = 2e^2/h$ ), which were reversibly opened and closed.

(b) Idealized geometries of silver point contacts with predefined numbers of contacting atoms. Conductance calculations performed within a Landauer approach result in near-integer multiples of $G_0$ for each of the five contact geometries (#1-5). For the conformations shown above, the axis of symmetry of the junction corresponds to the crystallographic [111] direction.

FIG. 3.

Calculations of the transmission as a function of the electron energy for the five different silver contacts (#1-5) of Fig. 2(b). The experimentally relevant values correspond to the conductance at the Fermi energy indicated by the vertical line in the figure.

Figure 1

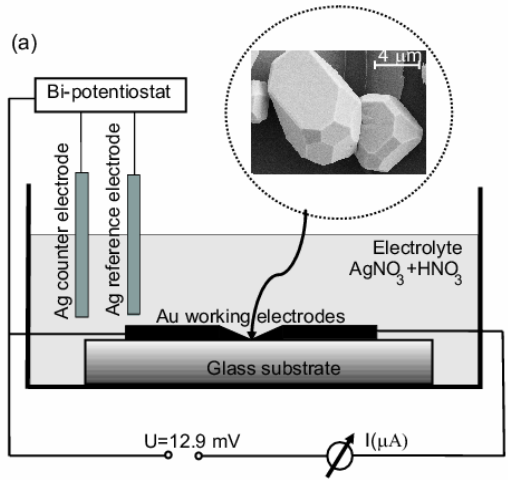

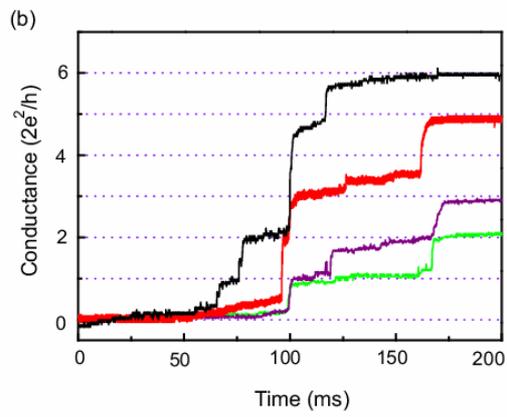

Figure 2

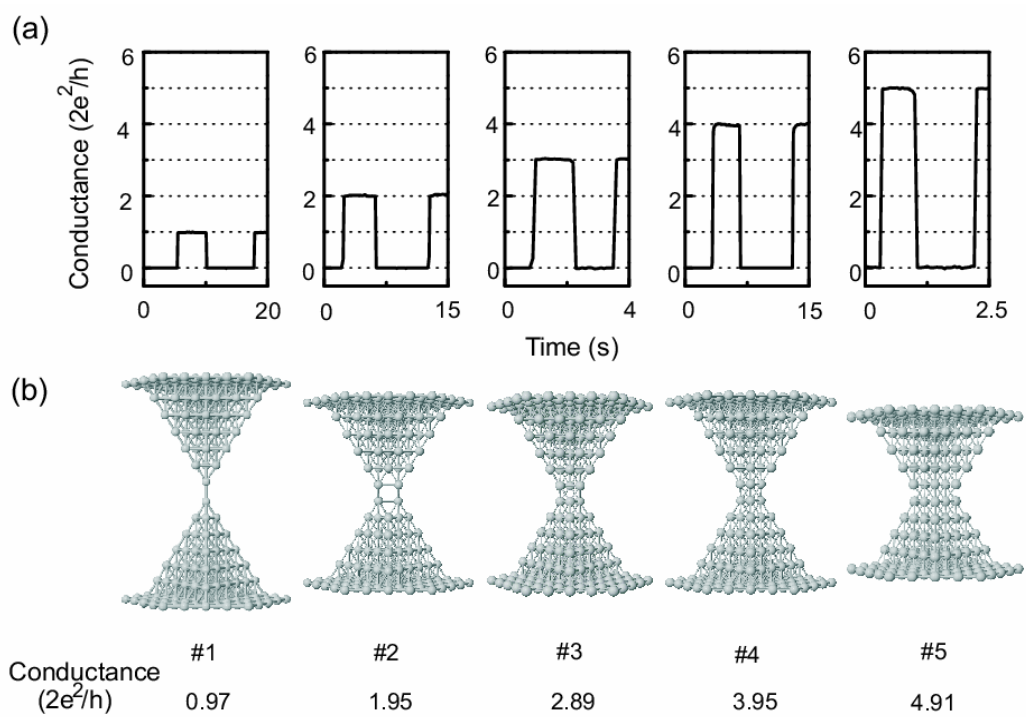

Figure 3

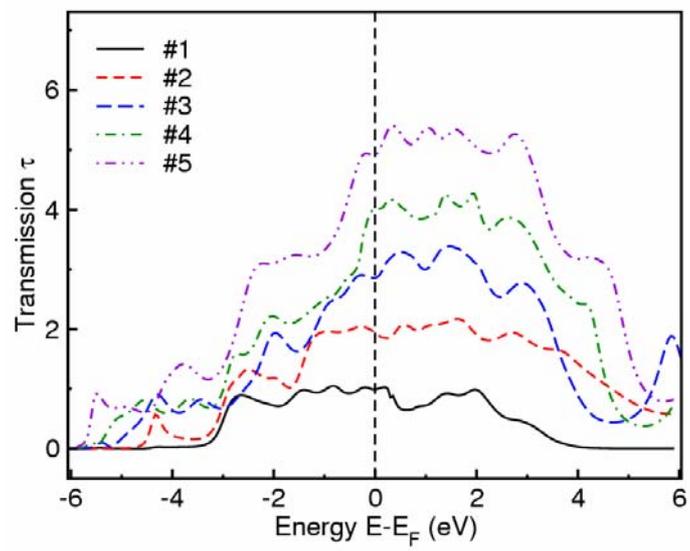